\documentstyle[psfig,aas2pp4]{article}
\lefthead{Henriksen}
\righthead{}

\begin{document}

\title{RXTE and ASCA Constraints on Non-thermal Emission from the
 A2256 Galaxy Cluster}

\author{Mark Henriksen}
\affil{Physics Department, University of North Dakota,
    Grand Forks, ND 58202-7129}
\centerline{mahenrik@plains.NoDak.edu}

\begin{abstract}
An 8.3 hour observation of the Abell 2256 galaxy
cluster using the Rossi X-ray Timing Explorer (RXTE) 
proportional
counter array (PCA) produced a high quality
spectrum in the 2 - 30 keV range. Joint fitting with the 0.7 - 11
keV spectrum obtained with the Advanced Satellite for Astrophysics
and Cosmology (ASCA) Gas Imaging spectrometer
(GIS) gives an upperlimit
of $\sim$2.3$\times$10$^{-7}$ photons cm$^{-2}$
sec$^{-1}$ keV$^{-1}$ for non-thermal emission at
30 keV. This yields a lower limit to the mean magnetic
field of 0.36$\mu$G and an 
upperlimit of 1.8$\times$10$^{-13}$ ergs cm$^{-3}$
for the cosmic-ray electron energy density. The resulting
lower limit to the central magnetic field is $\sim$1 - 3 $\mu$G. 
While a magnetic field of $\sim$0.1 - 0.2 $\mu$G can be
created by galaxy wakes,
a magnetic field of several $\mu$G is usually associated with a cooling
flow or, as in the case of the Coma cluster, a subcluster merger.
However, for A2256, the evidence for a merger is weak
and the main cluster shows no evidence of a cooling flow. Thus,
there is presently no satisfactory hypothesis for the origin of
an average cluster magnetic field as high as $>$ 0.36 $\mu$G
in the A2256 cluster.
\end{abstract}

\section
{Introduction}
The Abell 2256 galaxy cluster ($z$ = 0.058), like the
Coma cluster, is one of the 
most studied galaxy clusters.
A2256 and the Coma cluster share many common properties: they have similar
X-ray luminosities, both have
optical and X-ray substructure, and both have a radio halo.
As with Coma, it has been over 20 years since the radio halo of A2256 
was discovered
(Bridle \& Fomalont 1976; Bridle et al. 1979).
These authors report that the radio halo has a
flux of 0.1 Jy at 610 MHz and extends to a radius of 5 arcmin
or 0.52H$_{o}^{-1}$ Mpc. The spectrum
is a powerlaw given by, S = 6.5$\times$10$^{-9}$$\nu^{-1.8}$
ergs cm$^{-2}$ sec$^{-1}$Hz$^{-1}$.The spectral index
is determined between 610 and 1415 MHz. As with Coma, 
the radio spectrum is attributed
to synchrotron emission from a population of relativistic
electrons interacting with a diffuse intergalatic magnetic
field. 
Recent studies of the Coma cluster link the X-ray
and radio emission to support the hypothesis that the
merger amplifies the magnetic field. 
However, for A2256 this
hypothesis may not work since the evidence for a merger
is incomplete; there is evidence of substructure, yet shock
heating due to the merger is not apparent in the temperature
map obtained with {\it ASCA} (Markevitch \& Vikhlinin 1997)
as it is for Coma.

X-ray emission is also expected from inverse-Compton
scattering of cosmic microwave photons off of the relativistic
electrons which produce the radio halo. So far, attempts to detect non-thermal
emission have only produced upper limits for both clusters. Previous searches
for non-thermal emission from A2256 have 
utilyzed hard X-ray data obtained with
HEAO1-A4 (Rephaeli, Gruber, \& Rothschild 1987). 
An upperlimit
 to the non-thermal X-ray flux reported from this study is
5$\times$10$^{-6}$ photons cm$^{-2}$
sec$^{-1}$ keV$^{-1}$ at
30 keV. The lower limit to the average magnetic field of A2256
is 0.11 $\mu$G.

The alternative hypothesis to mergers, that 
galaxy wakes amplify a seed field,
leads to an average field in the
range of 0.1 - 0.2 $\mu$G (Goldman \& Rephaeli 1991). 
If this hypothesis is correct, non-thermal
emission should be detectable for A2256 with even a slight
improvement in sensitivity over the HEAO-1 A4 X-ray spectrum.
In this paper, we use the
very high signal to noise data
obtained with the {\it ASCA} GIS combined with 
that obtained with the {\it RXTE} PCA
to place a more sensitive limit
on the non-thermal X-ray emission from the A2256 cluster. This
new limit provides a more stringent constraint on current
hypotheses for its origin.
A value of 50 km s$^{-1}$ Mpc$^{-1}$ is used throught this
paper.

\section{Observations}

The Abell 2256 cluster was observed with {\it RXTE} on 1997 June 24-26
for 30,000 seconds. Data obtained with the
PCA is analyzed and presented here. The PCA consists of five
proportional counter units (PCU), each of which contains
three detector layers. The nominal energy band
of the PCA is 2 - 60 keV. After filtering the PCA
data, there were $\sim$22,000 seconds
of good
data.
Filtering involved excluding data when less than 5 PCUs
were on (this eliminated $<$1000 seconds), excluding 
data taken when the pointing was $>$0.02 degrees offset
from the source, excluding data taken $<$10 degrees 
elevation from the Earth, exlcuding times 
near passage through the South Atlantic Anomaly which
showed large variations in the count rate. 
The observation was carried out in two
observing periods, each resulting in $\sim$11,000
seconds of filtered data. 
Three {\it RXTE} spectra were modeled:
that obtained for the top layer
detectors alone for each observation period, both separate
and combined, 
and the top and 
middle layer detectors. 
The background was modeled using the latest
models available from the {\it RXTE} Guest Observer Facility. 
The background estimation is based on three models:
VLE, activation, and cosmic X-ray background. The VLE
(based on the rate of Very Large Event discriminator)
background results from interaction of cosmic-rays with the
spacecraft or detector and unvetoed cosmic-ray tracks. 
The activation model depends on orbital
coordinates and corrects for 
additional background associated with passes from the
South Atlantic Anomaly. 
The cosmic X-ray background component is derived from blank sky
observations taken in 1997 to give an average high-latitude
X-ray background spectrum.

The model background 
consists of the VLE and activation
components (in which sky background is included from
blank sky observations). 
In the third layer, in the 30 - 60 keV range,
little source emission should be present. Indeed,
the count rate in the 2 - 30 keV 
range between the first and second layer
drops by a factor of $\sim$15 and the $\sim$7 keV
thermal continuum which characterizes the
A2256 X-ray spectrum contributes little emission above 30 keV.
The model background is
fit to the 30 - 60 keV spectrum, while normalizing the activation
component separately to minimize $\chi^{2}$.

The 2 - 30 keV
part of the spectrum used in the model fitting
utilyzes the background derived by this procedure.
The best contraint on the 
non-thermal flux was obtained fitting
data obtained in the top and second layers in the first
11 ksec observing period. The background subtracted count rates for the
data analyzed are:
28.43$\pm$0.23 counts sec$^{-1}$ in the top layer of the
PCA, 1.91$\pm$0.07 counts sec$^{-1}$ in the second
layer, and 1.16$\pm$0.05 counts sec$^{-1}$ in the third layer.

The spectral response of the PCA depends on 
the overall gain setting, which PCUs are analyzed
since they have different gains, and the detector layer.
Separate response matrices were created for each
detector layer of each PCU used. The matrices were binned
to the 129 channel, standard 2 configuration in which
the data were analyzed. There is a feature in the
response from the detector due to the Xenon L 
edge at 4.78 keV which produces
and absorption feature in the spectrum. Ignoring channels
10 - 11 in the PCA greatly reduces the residuals
in this part of the spectrum; this is done is all of the
model fitting which involves the PCA.

The {\it ASCA} GIS spectrum consists of
34,726 seconds of good data with a count rate of 1.57$\pm$0.007
counts s$^{-1}$. Preparation of the GIS spectrum 
and related calibration issues are discused in
detail in Henriksen (1998). 

\section{Analysis and Results}

The models fit to the data consist of one
or two Raymond \&
Smith (RS) (1977) thermal components with and without a power law
component.
For the thermal components,
all data groups 
share the following free parameters: column density,
temperature, and abundance. The normalization(s) for the PCA data sets
are tied while the GIS has
its own normalization(s). 
The powerlaw component has a free normalization and the spectral
index is fixed at the photon index measured
from the radio, 2.8, which is also the spectral
index of the inverse-Compton component. 
Both the GIS and the PCA field of
view contain the region of the radio halo so their
powerlaw 
normalizations are tied. Thus, for the joint data
fits, there are 6 free
parameters for the single RS component plus a
powerlaw and 8 free parameters for the model
consisting of 2 RS components and a powerlaw.
The 90\% range on fit parameters are given
in Tables 1 and 2 for each data set alone and the joint
fit. 

\begin{deluxetable}{ccccccc}
\footnotesize
\tablewidth{0pt}
\tablecaption{Results of Single and Joint Fits}
\tablehead{
\colhead{Model} & \colhead{Data Set} & \colhead{n$_{H}\times10^{22}$cm$^{-2}$} & \colhead{kT(keV)} &
\colhead{Abundance} & NT Normalization\tablenotemark{a}  & $\chi^{2}$/dof 
}
\startdata

1RS			& GIS 	& 0.028 - 0.061				& 6.78 - 7.44 	& 0.16 - 0.24	& $<$0.0045 	& 633.0/604	 \nl
1RS     		& PCA		& 0.042\tablenotemark{b}	& 6.63 - 6.81	& 0.15 - 0.18    & -			& 172.6/114	\nl
1RS + POW		& PCA		& 0.042\tablenotemark{b}      	& 7.10 - 7.43	& 0.23 - 0.27 	& 0.0091 - 0.015  & 127.6/113	 \nl
1RS			& Joint 	& 0.042 - 0.066			& 6.66 - 6.84	& 0.16 - 0.18	& -			& 815.8/720	 \nl
1RS + POW		& Joint	& 0.091 - 0.18				& 6.79 - 7.08	& 0.18 - 0.22	& 0.0019 - 0.0087 	& 806.7/719  \nl
2RS			& PCA		& 0.042\tablenotemark{b}	& 6.97 - 7.59	& 0.18 - 0.24  	& $<$0.0082		& 122.5/113	\nl
			& 		& 				&  0.71 - 1.71	&		& 			& 		\nl
2RS			& Joint 	& 0.029 - 0.11			& 6.99 - 7.38	& 0.19 - 0.23	& $<$0.0032		& 756.0/717 \nl
			& 	 	& 					& 0.75 - 1.46	& 		 	&			&		\nl
\enddata
\tablenotetext{a}{Non-thermal upperlimit in photons cm$^{-2}$s$^{-1}$keV$^{-1}$ at 1 keV}
\tablenotetext{b}{n$_{H}$ fixed at the Galactic value, 0.042}
\end{deluxetable}

\begin{deluxetable}{cccccc}
\footnotesize
\tablewidth{0pt}
\tablecaption{Results of Joint Fits}
\tablehead{
\colhead{Model} & \colhead{Data Set} & \colhead{Norm: High T} & 
\colhead{Norm: Low T} & \colhead{Flux: High T\tablenotemark{a}} 
& \colhead{Flux: Low T\tablenotemark{a}}
}
\startdata

1RS		& GIS 	& 0.087 - 0.089	& -			&  7.81		&  -	 	\nl
 -		& PCA 	& 0.071 - 0.073	& -			&  6.36		&  -	 	\nl
1RS + POW 	& GIS		& 0.078 - 0.085	& -			&  6.89		&  -		\nl 
-		& PCA		& 0.063 - 0.066	& -			&  5.78		&  -		\nl
2RS		& GIS 	& 0.084 - 0.087	& 0.00 - 0.011	&  7.97		& 0.00 	\nl
 - 		& PCA 	& 0.063 - 0.068 	& 0.024 - 0.077	&  6.12		& 0.31	\nl
\enddata 
\tablenotetext{a}{$\times$10$^{-11}$ergs cm$^{-2}$s$^{-1}$keV$^{-1}$ in 2 - 10 keV band. Flux is derived from
best fitting normalization.}
\end{deluxetable}

When addition of a powerlaw improves $\chi^{2}$,
the fit is also shown without the powerlaw to make the
significance of the additional component apparent.

The addition of a second temperature component is
significant using the PCA data alone as well as in the
joint fit which uses both the PCA and GIS data.
This is not a surprising feature of the integrated
spectrum since a temperature map of the
cluster shows multiple components (Markevitch \&
Vikhlinin 1997; Miyaji et al. 1993). 
The addition of the second, cooler
component provides much improvement in the joint fit; $\Delta$$\chi^{2}$
decreases by 60 for 3 additional degrees of freedom. 
The parameters derived
from fitting the GIS data and the PCA data separately
are in very good agreement, suggesting the
data are independently well calibrated. 
The PCA data alone do not constrain the Galactic
column density so this parameter is tied at the value
measured from neutral hydrogen, 4.2$\times$10$^{20}$.

The PCA data alone
is consistent with a detection of non-thermal emission when a
powerlaw is added to a single thermal component. However,
addition of a second thermal component rather than the
powerlaw provides an even better fit to the data. Addition
of a powerlaw to the two thermal components then degrades the
fit and the data provide an upper limit on non-thermal emission.
The joint fit with the GIS also shows that a second thermal component
provides a much better fit than does a powerlaw. The joint fit
of a 2 RS model also then gives the best upper limit on non-thermal
emission. The upperlimit on the non-thermal flux obtained from the
joint fit, 2.64$\times$10$^{-12}$ ergs cm$^{-2}$ s$^{-1}$ in the 2 - 10
keV band is $\sim$24 times better than that obtained by {\it HEAO1-A4}.
The best fit two RS model with a powerlaw set at the 90\% upperlimit
is shown in Figure 1.


The {\it RXTE} PCA temperature alone is 6.63 - 6.81 keV, in good agreement
with the {\it ASCA} GIS, 6.78 - 7.44 keV, and
measurements by previous X-ray 
observations: the {\it Einstein} MPC ( 6.7 - 8.1 keV;
David et al. 1993), and
{\it GINGA} (7.32 - 7.70 keV; Hatsukade 1989).
The abundance obtained from {\it RXTE}, 0.15 - 0.18 is consistent with
{\it ASCA}, 0.16 - 0.24. This general agreement with past measurements
suggests that the spectral calibration of the PCA is accurate
within the energy band used here.


\section{Calculation of $<B>$ and $U_{r}$}

The average magnetic field, $<B>$, is calculated from the
radio spectrum and the X-ray flux 
upperlimit using
the equations in Henriksen (1998). 
This procedure combines the expressions for the
synchrotron flux and the Compton
flux to eliminate the relativistic electron density to obtain 
an expression which is independent of the size of the emitting
region or the distance to the cluster. The derived values
of $<B>$ are given in Table 3.

\begin{deluxetable}{ccccccc}
\footnotesize
\tablewidth{0pt}
\tablecaption{Calculated Parameters}
\tablehead{
\colhead{Model} & \colhead{Data Set} & \colhead{NT Flux\tablenotemark{a}}  
& \colhead{$<$B$>$$\mu$G} & \colhead{U$_{r}$\tablenotemark{b}} 
& \colhead{U$_{B}$\tablenotemark{c}} 
}
\startdata

1RS	& GIS 	& $<$3.64 		& $>$0.30	& $<$2.55 	& $>$3.58	 \nl
	& PCA 	& 7.5 - 12.5	& 0.20 - 0.25& 5.3 - 8.6 & 1.6 - 2.5	  			\nl
	& Joint 	& 1.58 - 7.09	& 0.25 - 0.43 & 1.1 - 5.0 & 2.5 - 7.4		\nl
2RS	& PCA		& $<$6.80		& $>$0.25 & $<$4.8& $>$2.5		\nl
	& Joint     & $<$2.64		& $>$0.36 & $<$1.8 & $>$5.2 	\nl
\enddata
\tablenotetext{a}{Non-thermal flux: $\times$10$^{-12}$ergs cm$^{-2}$s$^{-1}$ in 2 - 10 keV band}
\tablenotetext{b}{Electron energy density: $\times$10$^{-13}$ergs cm$^{-3}$}
\tablenotetext{c}{Magnetic field energy density: $\times$10$^{-15}$ergs cm$^{-3}$}
\end{deluxetable}

The highest lower limit is 0.36$ \mu$G.
An upperlimit to the
energy density of relativistic electrons is calculated using
the following equation:
\begin{equation}
U_{r} = C\int(\gamma E)\gamma^{-p}d\gamma.
\end{equation}
In this equation, $U_{r}$ is the relativistic
electron energy density, $\gamma$ is the Lorentz factor,
$E$ is the electron rest energy, 
and $C$ and $p$ are defined by the
relativistic electron density distribution,
\begin{equation}
n(\gamma) = C\gamma^{-p}.
\end{equation}
The lower limit on the integral is the minimum $\gamma$ 
needed to boost a CMBR photon to the X-ray regime, 1000.
Solving the integral gives,
\begin{equation}
U_{r} = \frac{CE\gamma_{min}^{2-p}}{p-2}.
\end{equation}
The spectral index of the relativistic electrons ($p$)
is related to the photon index used to model the inverse-Compton 
emission ($\alpha_x$ = 2.8)
by $p$ = 2$\alpha_x$ - 1.
The parameter $C$ is calculated by multiplying equation
(4) in Henriksen (1998), which is the expression for the
inverse-Compton energy density in terms of various constants,
by the volume containing the relativistic electrons
which is taken to be a sphere with radius equal to that of the radio halo. 
This equation is then divided by 4$\pi D^{2}$, where $D$ is
the distance to the cluster to get 
the inverse-Compton flux. The inverse-Compton flux,
given in Table 3 in the 2 - 10 keV energy band is the primary observational
constraint in this calculation. The calculated
upper limits on $U_{r}$ are given in Table 3. The
magnetic field energy density is calculated for
comparison and shown in Table 3. In all cases, the
energy density in cosmic rays and the magnetic field
are consistent with equipartition.

\section{Discussion}

A joint fit of the GIS and PCA data in the 0.7 - 30 keV
band show that the A2256 spectrum is best described 
by a model consisting of 2 thermal components. Using a single
thermal component leads to a detection of non-thermal emission.
However, a second thermal component provides a better
description of the data and obviates
the need for non-thermal emission. These data place a
more stringent upper limit on 
the flux of non-thermal emission, $<$2.64$\times$10$^{-12}$
ergs cm$^{-2}$ s$^{-1}$ in the 2 - 10 keV band, or $<$4\% of the
thermal component. From the upper limit on non-thermal emission,
a lower limit to the average magnetic field is calculated, $>$0.36 $\mu$G.
If the field is assumed to be frozen into the intracluster gas,
then the central field will be higher. Using the
gas density parameters for A2256 
(Jones and Forman 1984) and the formulae in
Goldshmidt and Rephaeli (1994), a central field of $\ga$1 - 3 $\mu$G
is calculated. The range corresponds to using the radio
halo or the intracluster medium for the average magnetic field extent.
Using these parameters (central gas density = 2.5$\times$10$^{-3}$ cm$^{-3}$)
and the ambient gas
temperature of 7 keV gives a central gas energy density of 3$\times$10$^{-11}$
ergs cm$^{-3}$. The highest lower limit to the magnetic field
energy density at the center, $>$3.6$\times$10$^{-13}$, is
almost a factor of $\sim$83 lower. The energy density in cosmic-ray electrons,
$<$1.8$\times$10$^{-13}$, is also much lower. 
This gives an indication of the relative importance of the
magnetic field and cosmic-ray electrons on the gas dynamics
of the cluster. The cosmic-ray energy is not significant. 
The magnetic field energy is a lower limit and not
constraining. Detection of an average magnetic
field which gives a comparable energy density to the gas
at the center would require an increase in sensitivity 
in X-ray flux of $\sim$470.

The hypothesis that 
galaxy wakes amplify a seed field
via the dynamo process leads to an
average field of 0.1 - 0.2 $\mu$G (Goldman \& Rephaeli 1991).
These authors use a lower, more realistic efficiency for the 
conversion of kinetic energy into magnetic field
energy than earlier studies which produced a field of $\sim$2 $\mu$G
(Ruzmaikin, Sokoloff, \& Shukarov 1989).
Amplification for seed fields has also been considered
by De Young (1992) using time-dependent evolution of magnetohydrodynamic
turbulence from galaxy wakes. He found that amplification to
a few $\mu$G with this mechanism is very rare.
Magnetic field strengths of
a few $\mu$G have previously been inferred from
studies of the excess rotation
measure (RM) (Kim et al. 1990; Crusius-Watzel et al. 1990;
Kim, Kronberg, \& Tribble 1991) in galaxy clusters. 
Though these high magnetic 
fields challenge the hypothesis of wake amplification,
they are more model dependent as
they are sensitive to local density
inhomogeneities or an asymmetric gas distribution. Clusters often have 
both due to accretion of groups and subcluster mergers.
In addition, many other clusters also have 
cooling flows which can amplify the magnetic
field through isotropic compression (Soker \& Sarazin 1990; Tribble 1991).
The distribution of excess rotation measures
for Abell clusters found by Kim et al. is consistent
with amplification by galaxy
wakes if 
$\sim$10\% have cooling flows which further
amplify the magnetic fields (Carvalho 1994). However,
this requires that clusters with the larger 
magnetic fields, $>$0.1 - 0.2 $\mu$G, be those
clusters with cooling flows. Based on a deprojection
analysis of it's X-ray emission,
A2256 does not have a cooling flow (Stewart
et al. 1984) and does not show the surface brightness
central excess which is typical of clusters
with cooling flows (Jones \& Forman 1984).

As an
alternative, Tribble (1993)
has suggested that cluster mergers have amplified the
magnetic fields in clusters with radio halos. 
This hypothesis
has received support in subsequent studies of the radio
halo in Coma which suggest that
it may be maintained by the recent merger experienced
in the cluster center which is apparent in the X-ray (Deiss
et al. 1997). Bohringer et al. (1992) find that the
available energy specifically from the accretion of groups
near the core may be enough to reaccelerate cosmic-ray 
electrons and amplify the magnetic field to maintain
the halo. However, this hypothesis presents a problem in the
case of A2256 because the evidence for a merger is controversial.
The pattern of heating expected in a merger has been found in several
of the clusters (Coma, A754, and A1367). 
Non-isothermality in the atmosphere of A2256 was reported using
{\it ROSAT} PSPC data (Briel \& Henry 1994), {\it BBXRT} data 
(Miyaji et al. 1993),
and {\it ASCA} Markevitch \& Vikhlinin (1997). However, the latter
author's
found that the temperature
map was consistent with the superposition of
two subclusters with different temperatures rather than
a merger.
Future observation of A2256 by the Advanced X-ray Astrophysics Facility (AXAF)
will likely provide a more definitive description of the dynamical
state of the cluster. Thus,
a satisfactory hypothesis for the origin of
an average magnetic field $>$ 0.4 $\mu$Gauss
in A2256 is not yet apparent.



\acknowledgements

I thank the National Science Foundation for supporting this research. I
also thank the referee for comments on improving this manuscript.

\onecolumn

\psfig{figure=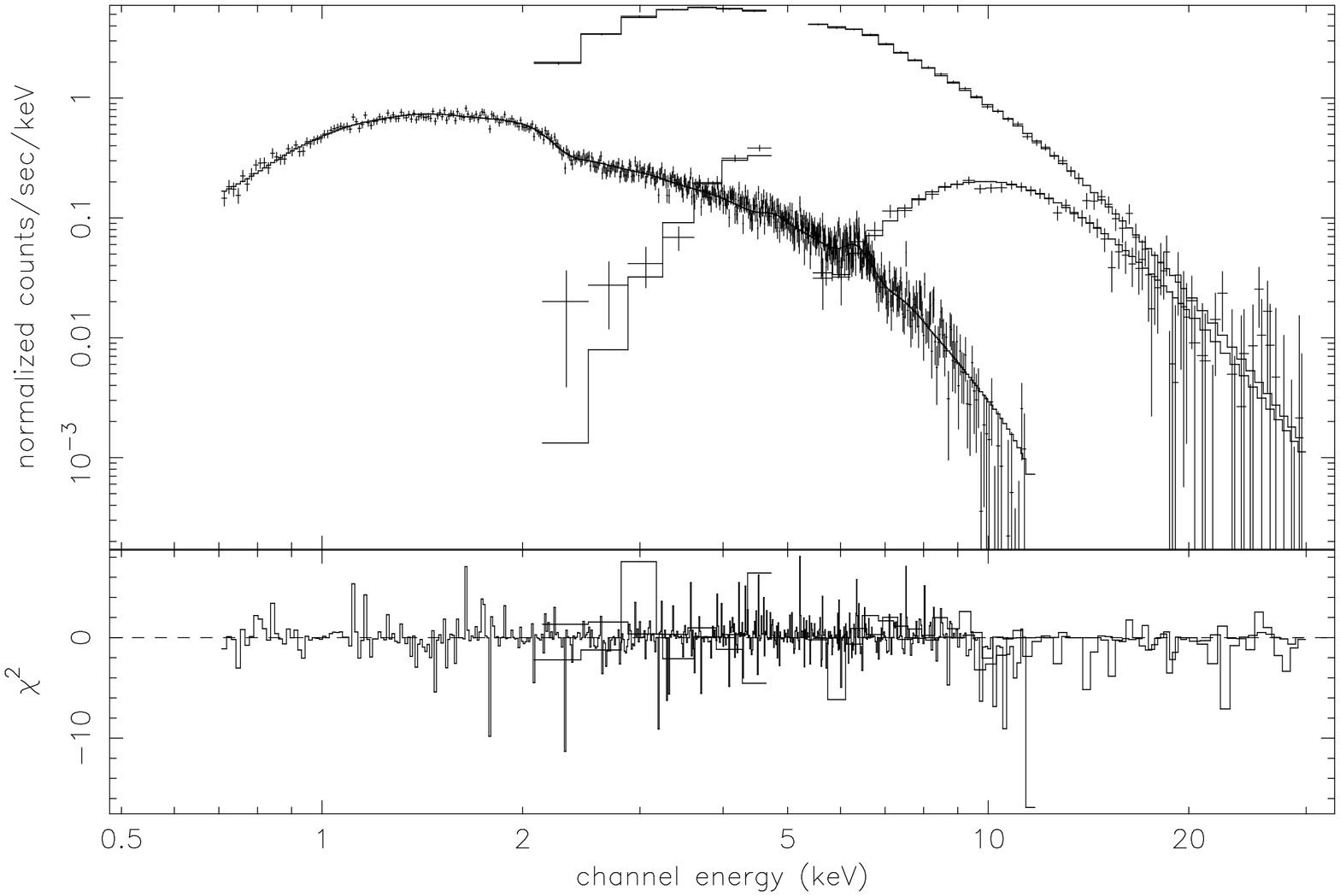,height=10cm,width=16cm,angle=-90}

\figcaption[]{Best fit model for joint fit to ASCA GIS and RXTE PCA
data shown with a powerlaw component added at the 90\% confidence 
upperlimit.}\label{fig1}


\begin{thebibliography}{}
\bibitem[Bohringer et al. 1992]{}Bohringer, H., Schwarz, R.,
 Briel, U, Voges, W., Ebeling, H.; Hartner, G.; Cruddace, R., 1992, in
  Clusters and Superclusters of Galaxies: ed. A. Fabian, (NATO/ASI), 71
\bibitem[Briel \& Henry 1994]{}Briel, U., and Henry, J., 1994, Nature, 372, 439
\bibitem[Bridle \& Fomalont 1976]{}Bridle, A., \& Fomalont, E., 1976, A\&A, 52, 107
\bibitem[Bridle et al. 1979]{}Bridle, A., Fomalont, E., Miley, G., Valentijn,
	1979, A\&A, 80, 201
\bibitem[Carvalho 1994]{}Carvalho, J.C. 1994, A\&A, 281, 641
\bibitem[Crusius-Watzel et al. 1990]{}Crusius-Watzel, A., Biermann, P., Lerche,
	I., \& Schlickeiser, R. 1990, \apj, 360, 417
\bibitem[David, et al. 1993]{}David, L.P., Slyz, A., Jones, C., Forman, W.
	Vrtilek, S.D., Arnaud, K.A., 1993, \apj, 412, 479.
\bibitem[Deiss et al. 1997]{}Deiss, B., Reich, W., Lesch, H., and Wielebinski,
	R., 1997, A\&A, 321, 55
\bibitem[De Young 1992]{} De Young, D., 1992, \apj, 386, 464
\bibitem[Goldman \& Rephaeli 1991]{}Goldman, I., Rephaeli, Y., 1991, \apj, 380
\bibitem[Goldshmidt \& Rephaeli 1994]{} Goldshmidt, 0., and
	Rephaeli, Y., 1994, \apj, 431, 586
\bibitem[Hatsukade 1989]{} Hatsukade, I., 1989, Ph.D. Thesis, Osaka University
\bibitem[Henriksen 1998]{}Henriksen, M., 1998, \pasj, 50, (in press)
\bibitem[Jones \& Forman 1984]{}Jones, C., and Forman, W., 1984, \apj,
	276, 38
\bibitem[Kim et al. 1990]{}Kim, K.-T., Kronberg, P.P., Dewdney, P.E.,
	Landecker, T.L., 1990, \apj, 355, 29
\bibitem[Kim, Kronberg, \& Tribble 1991]{} Kim, K. -T., Kronberg, P. P., Tribble, P. C., 1991, \apj, 379, 80
\bibitem[Markevitch \& Vikhlinin 1997]{}Markevitch, M., \& Vikhlinin, A., 1997,
	\apj, 474, 84
\bibitem[Miyaji et al. 1993]{}Miyaji, T., et al., 1993, \apj, 419, 66 
\bibitem[Raymond \& Smith 1977]{}Raymond, J., and Smith, B., 1977, \apjs, 35, 419
\bibitem[Rephaeli, Gruber, \& Rothschild 1987]{}Rephaeli, Y., Gruber, D., \&
	Rothschild, R., 1987, \apj, 320, 139
\bibitem[Ruzmaikin, Sokoloff, \& Shukarov 1989]{}Ruzmaikin, A., Sokoloff,
	D., Shukarov, A., 1989, \mnras, 241, 1
\bibitem[Soker \& Sarazin 1990]{}Soker, N., and Sarzin, C., 1990, \apj, 348, 73
\bibitem[Stewart et al. 1984]{} Stewart, G., Fabian, A., Jones,
	 C., Forman, W., 1984 \apj, 285 1
\bibitem[Tribble 1991]{} Tribble, P. 1991, \mnras, 248, 741
\bibitem[Tribble 1993]{} Tribble, P. 1993, \mnras, 263, 31
\end{thebibliography}
\end{document}